\documentclass[twocolumn,prb,aps]{revtex4}
\usepackage{graphicx}%
\usepackage{dcolumn}
\usepackage{amsmath}

\begin{document}
%
\newcommand{\K}{{\vec k}}



\title{Insulator-metal-insulator transition and selective spectral weight transfer in a
disordered strongly correlated system}

\author{P. Lombardo, R. Hayn}
\affiliation{Laboratoire Mat\'eriaux et Micro\'electronique
de Provence associ\'e au Centre National de la Recherche
Scientifique. UMR 6137. Universit\'e de Provence, France }
\author{G.I. Japaridze}
\affiliation{Georgian Academy of Sciences, Tamarashvili 6, 0177
Tbilisi, Georgia}
\date{\today}

\begin{abstract}
We investigate the metal insulator transitions at finite temperature
for the Hubbard model with diagonal alloy disorder. We solve the
dynamical mean field theory equations with the non crossing
approximation and we use the coherent potential approximation to
handle disorder. The excitation spectrum is given for various
correlation strength $U$ and disorder. Two successive metal
insulator transitions are observed at integer filling values as $U$
is increased. An important selective transfer of spectral weight
arises upon doping. The strong influence of the temperature on the
low energy dynamics is studied in details.
\end{abstract}


\maketitle


\section{Introduction}
Strongly correlated materials have been the subject of intensive
researches for the last forty years. The simplest lattice model
accounting for correlations is the one band Hubbard model. Great
progresses have been achieved within the last ten years with the
development of the dynamical mean field theory (DMFT) (for a review
see Ref.~\onlinecite{GKK96}). One important result of the DMFT is
the description of the metal to insulator transition driven by
correlations (Mott transition\cite{M90}) for the half-filled
Hubbard model. Today, the Mott metal insulator transition (MIT) is
the subject of intensive experimental and theoretical
studies.\cite{IFT98} Recently, a renew of interest for the Mott
transition has occurred in relation with the so-called orbital
selective Mott transition\cite{L03,L03_2,L05,KKR04} observed in
multi-orbital compounds.

In most of the experimental studies, strong electron correlations
are closely connected with disorder effects. So, to change
experimentally the band-filling, the band-width or other parameters,
one replaces usually one atom of a given compound by another one.
That is true for strongly correlated transition metal oxides or
similar compounds. For example, replacing La by Sr in
La$_x$Sr$_{1-x}$TiO$_3$ changes the doping, i.e.\ the
band-filling.\cite{FHNN92,HBTT03} In
Ca$_{1-x}$Sr$_x$VO$_3$,\cite{IHAF95,SFIS04} Ca and Sr atoms have
the same valence but different atomic radii, which changes the angle
of the V-O-V bond and, correspondingly, the band-width. More
recently, a very interesting study\cite{KLNL05} on the
metal-insulator transition in the disordered and correlated system
SrTi$_{1-x}$Ru$_x$O$_3$ has been performed. From their transport and
optical data, the authors propose a classification scheme with six
kinds of electronic states depending on $x$. The disorder and the
electron correlation effects should be considered together to
understand the measured electronic structure evolutions.

A generic model to study the common influence of disorder and
correlations is the Hubbard model including diagonal disorder. As we
will outline below, it shows many features and metal insulator
transitions not present in the standard Hubbard model. A possible
method to solve the disordered Hubbard model is the combination of
DMFT with the coherent potential approximation (CPA)\cite{G92} to
treat disorder. It was used in the seminal work of Laad {\it et
al.},\cite{LCM01} where the local impurity model of the DMFT was
solved by the iteration perturbation theory (IPT). Interesting
results concerning the stability of the Fermi liquid metal against
small disorder and the effect of the interplay between correlations
and disorder on the optical and Raman response had been obtained.

For the disordered Hubbard model, a first classification of the
metal-insulator transitions at {\em non-integer fillings} (and
$T=0$) was given in Ref.~\onlinecite{BHV04}. It used an analogy to
the Zaanen-Sawatzky-Allen scheme of charge-transfer or Mott-Hubbard
insulators, respectively (see also Ref.~\onlinecite{BHV05} for the
relationship to Anderson localization). Here, we will concentrate
first on the Hubbard model with diagonal alloy disorder at {\em
integer filling}. With increasing Hubbard correlation $U$ we find
two metal-insulator transitions: first from a non-correlated band
insulator to a metal and a second transition from a metal to a
Mott-Hubbard insulator.

The transition from a band insulator to a Mott-Hubbard insulator
has been the subject of intensive current studies within the
framework the half-filled
ionic Hubbard model (IHM).\cite{hub,nag} The IHM corresponds to the
spatially ordered limit of the binary alloy system which is
considered in the present paper, $A_{x}B_{1-x}$, where the $A$ atoms
are separated from each-other by an identical sequence of $B$ atoms.
In particular, intensive analytical and numerical efforts have been
performed to analyze the band insulator (BI) to Mott insulator (MI)
transition in the Hubbard chain with alternating on-site
energies.\cite{fab,tor,brune,man,poldi,otsu} These studies show that
the half-filled ionic Hubbard chain has two transitions as $U$ is
increased. The first is from the BI to a bond-ordered, spontaneously
dimerized, ferroelectric insulator (FI). The second, when $U$ is
further increased, involves a transition between the FI and the MI.
In this phase diagram the very unconventional metallic state is
realized only at the BI-FI transition point.

Very recent studies of the generalized $AB_{n-1}$ IH chain for $n
\geq 3$, where $n-1$ is the length of domain of $B$-atoms which
separates $A$ atoms, show that at commensurate band-filling $1/n$
the system shows, with increasing $U$, a similar sequence of
transitions BI-FI-MI.\cite{TAJN} Therefore, in the case of 1D
ordered binary chains, the BI phase is always separated from a MI by
the {\em insulating} ferroelectric phase. Although the Mott
transition in the disordered 1D binary chain has not been studied in
detail, the Anderson localization \cite{Anderson} excludes the
presence of a metallic phase in a finite sector of the ground state
phase diagram of a 1D disordered IH model.

Therefore, the approach presented in this paper allows us to
investigate in detail the band-insulator to correlated metal and
metal to Mott-Hubbard insulator transitions in a disordered
binary-alloy system for higher dimensions.\cite{Note-Randeria} Below
we analyze these metal-insulator transitions in detail. For that
purpose we consider the spectral weight transfer with doping, and we
find that it occurs in a selective way, different for the two
disorder sites in question.

In the present work, we propose a finite temperature calculation of
the dynamical properties by using the non-crossing approximation
(NCA)\cite{B87,PCJ93} to solve the DMFT local Anderson problem in
the self-consistent loop. Then we combine the DMFT with CPA which is
a very natural way since both methods use a local self-energy.
Spectral densities and spectral weight transfers will be computed
with the NCA. The role played by the temperature for the low energy
physics will be discussed in details.

In the following section the model Hamiltonian is presented. This
section also presents the application of the DMFT on this
Hamiltonian and a solving method based on the NCA. In the third
section, the main results are discussed. These are the
insulator-metal-insulator transition at {\em integer filling} and a
detailed analysis of possible transitions at {\em integer} and {\em
non-integer} fillings by using their spectral properties.

\section{Hamiltonian and solving method}

We consider the following Hamiltonian, where correlations and diagonal
disorder are present~:
\begin{eqnarray}
H=& &-t\sum_{<i,j>,\sigma} (c^+_{i\sigma} c_{j\sigma}+ H.c.)\nonumber\\
&+&\sum_{i,\sigma} \varepsilon_i n_{i\sigma}
+U\sum_{i}n_{i\uparrow}n_{i\downarrow} \; .
\label{hamiltonien}
\end{eqnarray}
In expression~(\ref{hamiltonien}) the sum $<i,j>$ is the sum over
nearest neighbor sites of a Bethe lattice. $c^+_{i\sigma}$
(respectively $c_{i\sigma}$) denotes the creation (respectively
annihilation) operator of an electron at the lattice site $i$ with
spin $\sigma$ and $n_{i\sigma}$ is the occupation number per spin.
$U$ is the on-site Coulomb repulsion and $t$ is the hopping term. A
fraction $1-x$ of sites (sites A) have a local on-site energy
$\varepsilon_i=\varepsilon_A$ and a fraction $x$ of sites (sites B)
have a local on-site energy $\varepsilon_i=\varepsilon_B$. We will
note $\Delta=\varepsilon_A-\varepsilon_B$ the energy difference
between the two types of sites.

Following the DMFT procedure of Ref.~\onlinecite{LCM01}, by integrating
out all fermionic degrees of freedom
except for a central site $i=o$, the lattice model~(\ref{hamiltonien})
can be mapped onto an effective impurity model.
The expression of the corresponding Hamiltonian $H^{A,B}_{\mathrm{eff}}$
depends on the nature of the central site $i=o$. It contains
a local part $H^{A,B}_{\mathrm{loc}}$ and a part corresponding
to the coupling to the effective medium. This coupling is determined
self-consistently. We then write~:
\begin{equation}
   H^{A,B}_{\mathrm{eff}}=H^{A,B}_{\mathrm{loc}}+H_{\mathrm{med}}~,
   \label{model}
\end{equation}
where
\begin{eqnarray*}
   H^{A,B}_{\mathrm{loc}}&=& U n_{o\uparrow}n_{o\downarrow} \nonumber\\
&+& \sum_{\sigma}(\varepsilon_{A,B}-\mu)c^+_{o\sigma}c_{o\sigma}~,
\end{eqnarray*}
is depending on the nature (A or B) of the central site $i=o$.
In the expression of $H^{A,B}_{\mathrm{loc}}$, we have written explicitly
the chemical potential $\mu$. The coupling Hamiltonian with the
effective medium is
\begin{eqnarray*}
   H_{\mathrm{med}}&=&\sum_{\K \sigma}
   \left(W_\K b^+_{\K \sigma} c_{o\sigma}+H.c.\right)\\
   &+&\sum_{\K \sigma}\varepsilon_\K b^+_{\K \sigma}b_{\K \sigma}~.
\end{eqnarray*}
%
$W_\K$ represents the
hybridization between the site $i=o$ and the
effective medium. $\varepsilon_\K$ is
the band energy of the effective medium. $b^+_{\K \sigma}$
(respectively $b_{\K \sigma}$) is the creation (respectively
annihilation) operator of an electron in the effective medium.
The effective medium can be
characterized by the effective dynamical hybridizations~:
$$
   {\cal J}(\omega)=\sum_{\K}\frac{|W_\K|^2}{\omega+i0^+-\varepsilon_\K}~.
$$
On a Bethe lattice, the self-consistent equations of the DMFT can be simply
written
$$
{\cal J}(\omega)=t^2 G_{\sigma}(\omega)~,
$$
where $G_{\sigma}(\omega)$ is obtained by averaging over disorder
with the CPA procedure. We have $G_{\sigma}(\omega)=x
G^B_{\sigma}(\omega) +(1-x) G^A_{\sigma}(\omega)$ where
$G^A_{\sigma}(\omega)$ (respectively $G^B_{\sigma}(\omega)$) is the
Green's function solution of the effective local impurity problem
represented by the Hamiltonian $H^{A}_{\mathrm{eff}}$ (respectively
$H^{B}_{\mathrm{eff}}$).

To solve the local impurity problem $H^{\alpha}_{\mathrm{loc}}$
(where $\alpha=A,B$) with the NCA, we introduce four
local states $|m^\alpha\rangle$ and their local energies $E^\alpha_m$.
$m=1$ corresponds to an empty local site, $m=2$ (respectively $m=3$)
corresponds to a singly occupied local site with an electron of
spin $\uparrow$ (respectively $\downarrow$), and $m=4$ corresponds
to a doubly occupied local site. Of course, energy $E^\alpha_1$ is
$\alpha$ independent and because of spin degeneracy, we have
$E^\alpha_2=E^\alpha_3=\varepsilon_\alpha-\mu$.
For the doubly occupied site $E^\alpha_4=2\varepsilon_\alpha-2\mu+U$.

The NCA equations\cite{B87} read~:
\begin{eqnarray}
   \Sigma^\alpha_1(\omega)&=&2\int d\varepsilon
   \tilde{\rho}(\varepsilon) f(\varepsilon)P^\alpha_2(\omega+\varepsilon)
   \nonumber\\
    \Sigma^\alpha_2(\omega)&=&\int d\varepsilon
   \tilde{\rho}(\varepsilon) f(-\varepsilon)P^\alpha_1(\omega-\varepsilon)
   \nonumber\\
   &&+\int d\varepsilon
   \tilde{\rho}(\varepsilon) f(\varepsilon)P^\alpha_4(\omega+\varepsilon)
   \nonumber\\
    \Sigma^\alpha_3(\omega)&=& \Sigma^\alpha_2(\omega)\nonumber\\
   \Sigma^\alpha_4(\omega)&=&2\int d\varepsilon
   \tilde{\rho}(\varepsilon) f(-\varepsilon)P^\alpha_2(\omega-\varepsilon)
    \nonumber
\end{eqnarray}
where the local propagators $P^\alpha_m$ are defined by~:
\begin{eqnarray*}
P^\alpha_m(\omega)
=\frac{1}{\omega+i0^+-E^\alpha_m-\Sigma^\alpha_m(\omega)}
\end{eqnarray*}
and the effective medium spectral density by $\tilde{\rho}(\varepsilon)=
-\frac{1}{\pi}{\mathrm{Im}}({\cal J}(\omega))$. $f$ is the Fermi-Dirac
distribution at temperature $T$.
We finally obtain the site specific one particle Green's function~:
\begin{eqnarray*}
G^\alpha_{\sigma}(\omega)&=&\frac{1}{Z^\alpha_0} \int d\varepsilon
{{e}^{-\beta\varepsilon }}[
p^\alpha_0(\varepsilon)P^\alpha_1(\omega+\varepsilon)+
p^\alpha_1(\varepsilon)P^\alpha_2(\omega+\varepsilon)\\
&-&p^\alpha_1(\varepsilon)P^\alpha_0(\varepsilon-\omega)^\star-
p^\alpha_2(\varepsilon)P^\alpha_1(\varepsilon-\omega)^\star
]
\end{eqnarray*}
where $p^\alpha_m(\varepsilon)=
-\frac{1}{\pi}{\mathrm{Im}}(P^\alpha_m(\varepsilon))$
and the partition function is $Z^\alpha_0=\int d\varepsilon
{{e}^{-\beta\varepsilon }}[
p^\alpha_0(\varepsilon)+2p^\alpha_1(\varepsilon)
+p^\alpha_2(\varepsilon)]~.$
%

\section{Results and discussion}
In this section, we present the calculated densities of states for
the disordered Hubbard model. In the first sub-section, we
focus on the possible metal insulator transitions and in
particular the nature of
these transitions.

In the second sub-section ~\ref{SSWT}, we show how the site selective
spectral weight transfer can help to understand the interplay
of disorder and correlations for dynamical properties.

\subsection{Metal insulator transitions}

One has to distinguish the metal insulator transitions at integer
filling ($n=1$) from those which occur at non-integer fillings
($n=x$ or $n=1-x$),\cite{BHV04} since at $\Delta=0$  the MI
transition with increasing on-site repulsion $U$ takes place only
in the former case.

\begin{figure}[hbtp]
\begin{center}\leavevmode
\includegraphics[width=8cm]{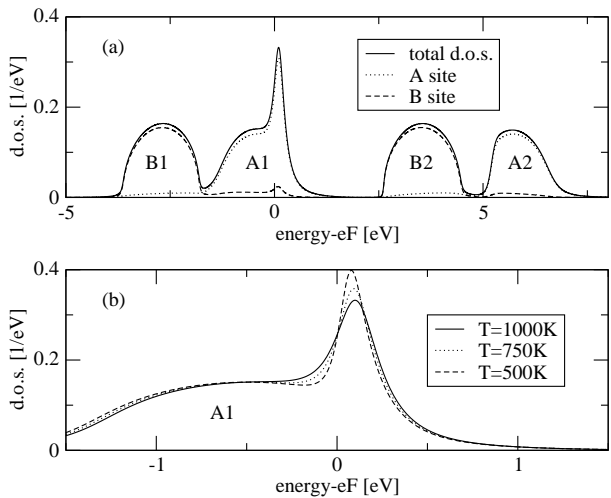}
\caption{(a) Density of states (plain line) obtained from DMFT-NCA and CPA for
$U=6$~eV, $\Delta=2$~eV and $T=1000$~K. Dotted line (respectively dashed line)
shows the corresponding density of states for the impurity problem with a site
A (respectively B) surrounded by the effective medium. (b) Temperature
dependence of the low energy part of the calculated spectrum.}
\label{dos_temp}
\end{center}
\end{figure}

Let us first concentrate on those with $n=1$. In that case, the
concentration of sites A has to be $x=0.5$. Then, for each
$\Delta$ and sufficiently large values of $U$ we observe an
insulating state for $n=1$. Before discussing the commensurate
$n=1$ case it is very instructive to consider the case of a doping
slightly below half-filling. In this case we have a strongly
correlated metal with heavy quasi particles for all $U$. That is
illustrated in Fig.\ \ref{dos_temp} for $\Delta=2$~eV, $U=6$~eV,
and $n=0.9$.

NCA applied to DMFT is
known\cite{PCJ93} to supply a fairly good solution even away from
half-filling. For this strongly correlated situation, we found
(plain line) a metallic state with four incoherent bands and one
coherent resonance at the Fermi level, witch is a signature of the
presence of quasiparticles. The temperature dependence of the
quasiparticle peak is shown in part (b) of the figure. Although the
density of states is almost constant at the Fermi level, the low
energy states near the Fermi level are strongly temperature
dependent. In part (a) of figure~\ref{dos_temp}, we plot as well the
spectral densities obtained from the solutions of the two impurity
models with a site A or B embedded in the same effective medium.
This gives enlightenment on the nature of the different bands
forming the density of states for the disordered system and then
allows to clarify the nature of the near insulating state. Using the
NCA, it is possible to reduce arbitrarily the doping with increasing
numerical efforts. Band A1 (respectively B1) is mainly composed of
singly occupied A sites (respectively B sites) and spectral weights
are proportional to the probability of removing one electron from a
site A (respectively B).
Correspondingly, the spectral weight of band A2 (respectively B2) is
proportional to the probability of adding one electron to a singly
occupied site A (respectively B).
The energy gap between band A1 and band A2 (or B1 and B2) is of the
order of $U=6$~eV. We can identify A1 to a {\it so called} lower
Hubbard band (LHB) and B2 to a {\it so called} upper Hubbard band
(UHB). We can therefore consider that the system is a strongly
correlated metal, close to a Mott insulator, with an insulating gap
between the LHB (A1) and the UHB (B2) and a reduced effective value
of the correlation strength $U-\Delta$. Note the asymmetric role
played by the two types of site. For $n=0.9$, charge fluctuations
are already blocked by correlations on sites B which are
half-filled. The Fermi level resonance peak is therefore only built
up by electrons evolving on sites A. This result is confirmed by the
respective local occupations $n_A=0.4$ and $n_B=0.5$.
%

\begin{figure}[hbtp]
\begin{center}\leavevmode
\includegraphics[width=8cm]{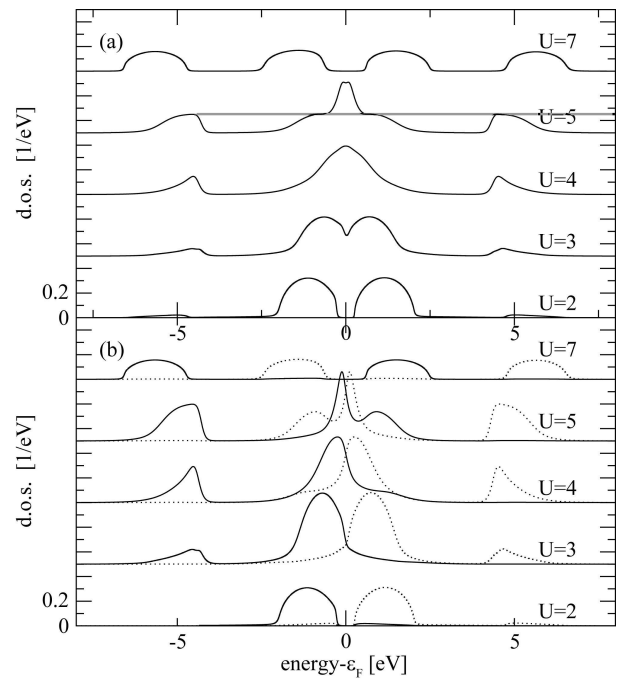}
\caption{(a) DMFT-NCA-CPA densities of states for $U=7,5,4,3,2$~eV,
$\Delta=4$~eV, $T=1000$~K, $x=0.5$ and $n=1.0$. (b) Corresponding
results for site selective densities of states.} \label{IMI_tr}
\end{center}
\end{figure}
%
Part (a) of Fig.~\ref{IMI_tr} shows the densities of states for a
decreasing $U$ ($U=7,5,4,3,2$~eV) and a given $\Delta=4$~eV at
$T=1000$~K for $x=0.5$ and $n=1$. We observe two phase transitions.
As expected, starting from the strongly correlated situation
$U=7$~eV, the system undergoes a transition to a metallic state for
decreasing $U$. There is a second critical value at which the system
enters into a new insulating state with further decreasing $U$. The
nature of the insulator is completely different here. This can be
clarified by examining part (b) of figure~\ref{dos_temp}. Sites B
are almost filled ($n_B=0.915$) whereas sites A are almost empty
($n_A=0.045$). The two side bands of the gap correspond to singly
occupied sites and cannot be interpreted in terms of Hubbard bands.
Therefore, this insulator can be understood as an ordinary band
insulator. The metallic state for intermediate values of $U$ can be
understood since in that case the energies for doubly occupied B
sites $\varepsilon_B+U$ and singly occupied A sites $\varepsilon_A$
are nearly degenerate. For $U=\Delta$ we have a special situation
(see Fig.~\ref{dos_temp}). In that case we have a single peak in the
DOS at the Fermi energy, whereas in all other situations a double
peak develops.

We believe that the series of phase transitions
insulator-metal-insulator is a rather generic situation and should
occur in a wide class of models with two different species at equal
concentration in dimensions higher then one. Very recently a similar
insulator-metal-insulator transition was obtained for the {\em
ordered} Hubbard model with two sites on a bipartite
lattice.\cite{Garg} Also, there is a quite large similarity to the
transition between a Mott-Hubbard insulator and a band-insulator
observed in the 1D ionic Hubbard model, however due to the
peculiarity of 1D systems the metallic state in this case is reduced
to only one critical point corresponding to the transition from a BI
to FI.

To characterize the different phases more in detail we computed also
the effective masses in a slightly doped situation ($n=0.96$, not
shown). We used the equation $m_\alpha^\star/m=1-\left[\partial
{\mathrm{Re}}\left({\Sigma^\alpha(\omega)}\right)
/\partial\omega\right]_{\omega=0}$. Crossing the upper
insulator-metal transition, we found a strong reduction of
$m_B^\star/m$ from $13.5$ (for $U=7$~eV) to $2.4$ (for $U=5$~eV)
while $m_A^\star/m$ evolves slightly from $1.5$ to $1.9$. This is in
contrast to the band-insulator case ($U=2$~eV) with the effective
masses $m_A^\star/m=1.004$ and $m_B^\star/m=1.07$ proving the
absence of heavy quasiparticles.

\begin{figure}[hbtp]
\begin{center}\leavevmode
\includegraphics[width=8cm]{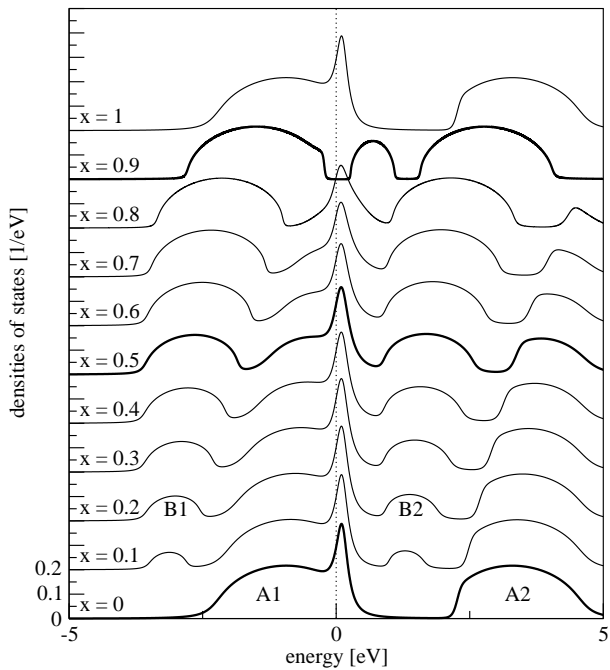}
\caption{Density of states obtained from DMFT-NCA and CPA for
$U=2\Delta=4$~eV, $n=0.9$ and various disorder concentrations
$x$ of site B between $0$ and $1$.}
\label{xvar}
\end{center}
\end{figure}

Finally, let us discuss the metal-insulator transitions at
non-integer fillings. They have been found by Byczuk~{\it et
al.}\cite{BHV04} Using the numerical renormalization group method
at zero temperature, they have shown that the system becomes a Mott
insulator at strong interaction for $n=x$ or $n=1-x$. Here we
propose a visual interpretation of this result. Figure~\ref{xvar}
displays the densities of states for $U=2\Delta=4$~eV, $n=0.9$ and
various disorder concentrations $x$ of site B between $0$ and $1$.
For $x=0$, there are only sites A whose spectral densities are
distributed over two Hubbard bands, and the lower one is partially
filled. Increasing the concentration of B sites leads to an
increasing strength of the two Hubbard peaks corresponding to B (B1
and B2). For $x=0.9$, the lower Hubbard peak of site B is completely
filled and sites A are nearly empty, leading to an insulating
situation. To observe the insulating state at $x=0.1$ we should
exchange the role of A and B site ($\Delta<0$).
%

\subsection{Selective spectral weight transfer}
\label{SSWT} 
\begin{figure}[hbtp]
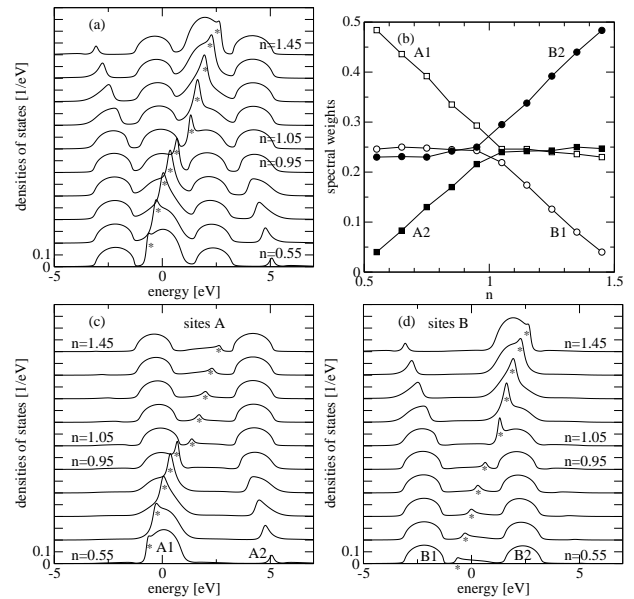

\begin{center}\leavevmode
\includegraphics[width=4cm]{transfer1.eps}
\includegraphics[width=4cm]{transferdat.eps}
\includegraphics[width=4cm]{transfer2.eps}
\includegraphics[width=4cm]{transfer3.eps}
\caption{(a) Densities of states for $U=\Delta=4$~eV, $T=1000$~K
and for various fillings from $n=0.55$ to $n=1.45$. (b) Spectral weights
of the different bands. (c) Corresponding densities of states for sites A.
(d) Corresponding densities of states for sites B.
The position of the Fermi level is marked with an asterisk
 on each spectrum.}
\label{transfer}
\end{center}
\end{figure}

In this section, we investigate a larger range of
doping values, from the quarter-filled situation where $n=0.5$ to
$n=1.5$, but keeping $x=0.5$. We found an important transfer of
spectral weight and explain it by using the selective spectral
densities supplied by our approach. Two particular cases of
non-integer filling $n=0.5$ and $n=1.5$ will be discussed in
details, with a new type of insulating state. In part (a) of
figure~\ref{transfer} the densities of states are displayed for
$U=\Delta=4$~eV, $T=1000$~K and for various fillings from $n=0.55$
to $n=1.45$. The position of the Fermi level is marked with an asterisk
on each spectrum. As the filling is increased, an important transfer
of spectral weight takes place between the different bands. The
spectral weights are reported in part (b) of the figure. Part (c)
(respectively part (d)) of the figure shows site selective densities
of states for sites A (respectively B). We found that the observed
spectral weight transfer is characteristic of strongly correlated
bands and is site specific. For $n$ going from $0.55$ to $0.95$ the
spectral weight transfer occurs for sites A (from $1.05$ to $1.45$
for sites B). For $n=0.55$ and $n=1.45$ the system is close to an
insulating state. Nevertheless, the nature of the insulating phase
is strongly different from the half-filled situation, where we have
a Mott insulator and two half-filled spectral densities for sites A
and B. For the almost quarter-filled situation ($n=0.55$) we found
an hybrid type of insulating phase, where sites B are half-filled
and sites A are almost empty. For B sites, charge fluctuations are
freezed by correlations like in the usual Mott insulator. For A
sites, the blocked microscopic processes are of charge transfer type
like in a band insulator.
For $n=1.45$ we have the symmetric situation.
%

\section{Conclusion}
In this paper, we proposed an approach based on the dynamical mean field
theory which is able to handle diagonal disorder in a strongly
correlated Hubbard model at finite temperature for any doping.
We used the coherent potential approximation in the self-consistent
equation of the DMFT and we applied the non-crossing approximation
to the local impurity problem on which the lattice model is mapped.

For a decreasing $U$ we showed that the disordered half-filled
system undergoes two successive metal insulator transitions of
different nature. The first transition is a Mott like transition
with a reduced effective correlation strength. The second
transition is a charge transfer like transition. We pointed out
that this situation is a very generic one. We also discussed the
differences and similarities between the considered disordered
binary-alloy model and the ionic Hubbard model in one and higher
spatial dimensions.

In addition to the insulating phase at half filling where both types
of site are half-filled and all charge fluctuations are blocked by
correlations effects, we found an hybrid type of insulating state
away from half filling, i.e.\ at $n=x$ or $n=1-x$. We illustrated it
for the quarter filled and the three-quarter filled situations.
Charge fluctuations are blocked by correlations for one type of
sites and by charge-transfer excitations for the other type of
sites.

It could be interesting to include in the present model
two orbitals with different bandwidth to investigate the role of disorder
in the orbital selective Mott transition.

We thank M. Laad and L. Craco for useful discussions and the NATO
science division for financial support (grant CLG 98 1255).
Authors would like to thank the generous
hospitality of the MPI-PKS Dresden, where their joint work on the
given problem had started.

\end{document}